\documentclass[12pt]{article}
\usepackage[utf8]{inputenc}
\usepackage{amsmath}
\usepackage{amsfonts}
\usepackage{amssymb}
\usepackage{bbm}
\usepackage{lipsum}       
\usepackage{xargs}            

\usepackage[colorinlistoftodos,prependcaption,textsize=tiny]{todonotes}
\newcommandx{\unsure}[2][1=]{\todo[linecolor=red,backgroundcolor=red!25,bordercolor=red,#1]{#2}}
\newcommandx{\change}[2][1=]{\todo[linecolor=blue,backgroundcolor=blue!25,bordercolor=blue,#1]{#2}}
\newcommandx{\Sinfo}[1]{\todo[backgroundcolor=red!25,bordercolor=red,noline]{S:#1}}
\newcommandx{\Minfo}[1]{\todo[backgroundcolor=blue!25,bordercolor=blue,noline]{M:#1}}
\newcommandx{\Rinfo}[1]{\todo[backgroundcolor=yellow!25,bordercolor=yellow,noline]{R:#1}}

\begin{document}

\title{Higher Spin Black Holes in Three Dimensions: 
Comments on Asymptotics and Regularity}
\author{M\'{a}ximo Ba\~{n}ados$^{a}$, Rodrigo Canto$^{a}$ and Stefan Theisen$^{b}$  \\
{\small $^a$Departamento de F\'{\i}sica, P. Universidad Cat\'{o}lica de Chile, Santiago 22, Chile}\\
{\small $^{b}$ 
Max-Planck-Institut f\"ur Gravitationsphysik (Albert-Einstein-Institut), 14476 Golm, Germany}}

\maketitle

\begin{abstract}

In the context of (2+1)--dimensional $SL(N,\mathbb{R})\times SL(N,\mathbb{R})$
Chern-Simons theory
we explore issues related to regularity and asymptotics on the solid torus, for 
stationary and circularly symmetric solutions. We display and solve all necessary 
conditions to ensure a regular metric and metric-like higher spin
fields. We prove that holonomy 
conditions are necessary but not sufficient conditions to ensure regularity, and that 
Hawking conditions do not necessarily follow from them. Finally we give a general proof 
that once the chemical potentials are turn on -- as demanded by regularity -- the 
asymptotics cannot be that of Brown-Henneaux. 

\end{abstract}

\section{Introduction}

Higher spin black holes in three dimensions have been analyzed in great detail 
recently  --- see \cite{Ammonetal} for a review of the relevant features ---
mostly in their Chern-Simons formulation, i.e. in terms of  
the properties of flat connections on a manifold with the topology of a 
solid torus.  
While a metric and higher spin fields 
can be introduced in terms of the connections, discussing 
properties of the black hole in terms of the metric encounters difficulties. 
This is because the symmetry group is now much larger than the diffeomorphisms group 
and this makes the definition of a horizon ambiguous.  

In particular, it has been known for a long time that for Chern-Simons black holes the 
radial coordinate can be gauged away and most of their properties can be studied from 
connections at a fixed radius \cite{BanadosBrotzOrtiz}. Nevertheless, at the end of the 
day one should be able to write spacetime fields, globally well defined on the manifold 
under consideration. The aim of this note is to discuss the issue of horizon 
regularity and asymptotics, introducing a radial coordinate in an appropriate way.  

As already alluded to, the use of the 
word `horizon' is tricky because in higher spin theories 
there is no global definition for it. It is a gauge dependent concept.  
Without attempting to solve this problem, 
what we do is to consider $sl(N,\mathbbm{R})$ gauge fields on a solid torus. 
For the case $N=2$ one knows that these fields can be interpreted as black holes. 
The question is whether extending to $N>2$, one can think of these solutions as black 
holes carrying some extra charges, and whether for small values of these charges, 
they can be seen as perturbations of the $N=2$ case. 
Finally, we are also interested 
in the question whether these solutions can be regarded as perturbations of 
AdS space for large $r$, consistently with regularity on the whole solid torus.  

The correct asymptotic conditions in terms of a radial coordinate were analyzed in 
detail in \cite{CampoleoniHenneaux}. Precise fall-off behaviour for all fields is 
described, ensuring the $W_3$ symmetry acting as extended higher spin 
diffeomorphisms. 
We shall prove here that there is an incompatibility between regularity and AdS 
asymptotics, as described in \cite{CampoleoniHenneaux}. In other words, the conditions 
spelled out there cannot be achieved simultaneously with 
regularity at the center of the solid torus, within the class of stationary and 
circularly symmetric solutions.   

Before plunging into details, we briefly summarize the content of this 
note. In the next section we carefully consider the issue of regularity 
at the center of the solid torus and how this restricts the near-horizon structure 
of the metric and higher spin fields.  
In Sec.~\ref{sec:holonomies} we analyze the general structure of solutions 
near the horizon and characterize them in terms of the eigenvalues of 
the holonomies. We explicitely solve the holonomy conditions by expressing the 
chemical potentials in terms of the spin-2 and spin-3 charges  
in Section \ref{sec:GSH}. The simplest regular solution with four
charges is constructed in Section \ref{sec:SRS}. The asymptotics of this 
solution changes if 
one switches off the spin-3 charge.  
In the last section we show that this a general feature:  
we give an argument which 
shows that regularity at the horizon clashes with AdS asymptotics of the solution. 
Most of our discussion is specific to $N=3$, but can be generalized to arbitrary $N$.  

\section{Regularity in the vicinity of the `horizon'}
\label{sec:reg}

We are concerned with $sl(N,\mathbbm R)$ gauge fields on the solid torus. 
We parametrize this manifold by three coordinates with ranges,
\begin{eqnarray}
0 \leq\! & t & \!< 2 \pi \\
0 \leq\! & r & \!<   \infty \\
0 \leq\! & \phi & \! < 2 \pi
\end{eqnarray} 
The Euclidean time coordinate $t$ is periodic and describes the contractible cycle. The 
surfaces $\phi$=constant are planes para\-metrized by  $r,t$, which are polar-like 
coordinates around the center of the solid torus. We use the convention where $r=0$ 
describes the center of the solid torus. We loosely call  this surface `the horizon' 
since for $N=2$ one recovers the 2+1 BTZ black hole \cite{BTZ}.  

Regularity of gauge fields is often expressed as a condition on holonomies over the 
contractible cycle in the (Euclidean) $t$-direction. This is a necessary but not a 
sufficient condition. Further restrictions come from local properties at the center of 
the solid torus. This problem exists irrespective of 
the presence of higher spin fields and is, for example, 
already encountered in the three-dimensional pure gravity theory. 

Riemann curvature technology, available for spin two, allows to analyze 
singularity/regularity questions related to the metric $g_{\mu\nu}$. But there is no 
analogous structure for higher spin fields. Therefore, in order to establish 
regularity/irregularity of these fields, we need to consider them directly. 

The basic idea 
is to express all components of a given field in a regular set of 
coordinates (e.g. Cartesian), and demand the components to be regular and well-defined 
(i.e. they should have unique values) \cite{Deser}. 
These conditions are not new to black hole physics. The new ingredient here are the 
higher spin fields. We shall analyze the conditions on the components of the
spin-3 field $g_{\mu\nu\rho}$ that lead to a sensible and regular solution. 
The generalization to higher spins $N>3$ is straightforward.    

Regularity of  $r^2 dt^2 + dr^2$ (with $0 \leq t < 2\pi$) 
is well-known. It is simply the metric on the Euclidean plane expressed in 
polar coordinates (which do not cover the whole plane because they are singular 
at the origin). 

Even though this combination is a regular rank-2 tensor, it is not often stressed 
that both $rdt$ and $dr$ are individually {\it singular} one-forms.
We explain this point in detail.
The extension to spin-3 is then straightforward. 
 
The transformation between Cartesian (regular) and polar (irregular) coordinates 
$r = \sqrt{x^2+y^2}$ and 
$t = \arctan(y/x)$ implies the relationship between the one-forms 
\begin{equation}\label{transxy}
dr = {xdx + ydy \over \sqrt{x^2 + y^2}}  \ \ \  \ \ \ \ \ \ \ 
dt = {xdy - ydx \over x^2 + y^2}   
\end{equation} 
While $r$ is well defined at the origin $x=y=0$, $t$ is not as $x/y$ is undefined there. 
Regularity of any tensor component requires that it is finite and that it has 
a unique value. 
Uniqueness forbids undetermined expressions of the form 
$0/0$. Clearly $dt$ violates the finiteness condition while 
$r dt$ violates the uniqueness condition. 
To see this very explicitely, 
approach the origin along a path with slope $\lambda$, i.e. set $y=\lambda\, x$. 
This reveals that $r\,dt$ is finite but it depends on $\lambda$. The same is true 
for $dr$ and, for that matter, for any $0/0$ expression.\footnote{As a byproduct 
of this analysis consider the 
Reissner-Nordstrom black hole in $d=4$ with metric and gauge field 
(near the horizon at $r=r_0$)
\begin{eqnarray}
ds^2 &=& \left(1 - {r \over r_0} \right) dt^2 + {dr^2 \over 1 - {r \over r_0} } 
+ r^2 d\Omega \nonumber\\
 A &=& \left( {q \over r} - {q \over r_0}\right)dt
\end{eqnarray} 
The constant piece in the gauge field is added in order to have $A(r=r_0)=0$. 
One may naively conclude that  the coefficient of $dt$ in $A$ must vanish linearly 
to have regularity. But one must be careful with the nature of coordinates. 
The Schwarzschild coordinate $r$ is not a proper coordinate. The proper radial 
coordinate $\rho$ for this metric is related to $r$ by $\rho = 2\sqrt{r-r_0}$. Thus, 
when writing the solution in terms of the proper coordinate one finds the gauge field 
$A  \sim \rho^2 dt $, which is indeed regular (while $\rho\,dt$ is not).}.

In the same way it is straightforward to see that 
$ds^2=\alpha^2 r^2 dt^2 + dr^2$, with $\alpha$ a constant, is singular except for
$\alpha=1$. It is flat away from the origin where it has a conical singularity
if $\alpha\neq 1$. 
If $\alpha=n$ is an integer one finds an excess of $2\pi(n-1)$. 
A vector parallel 
transported around the origin comes back to itself, and the geometry
appears to be regular. This is however not the case. This is one example 
(we shall see others) where topological conditions are not sufficient for regularity. 
They are merely necessary conditions. 

Turning now to the spin-3 field, the most general Ansatz which guarantees
finiteness at the origin $r=0$ is  
\begin{eqnarray}\label{gens3}
ds^3_3 &=& f_1\, r^3 dt^3 + f_2\, dr^3 + f_3\, d\phi^3 + f_4\, r^2 dt^2 dr 
+ f_5\, r^2 dt^2 d\phi + f_6\, dr^2 r dt \nonumber\\ 
&& \ \ \ \  +f_7\, dr^2 d\phi+  f_8\, d\phi^2 rdt + f_9\, d\phi^2 dr+f_{10}\, rdtdrd\phi
\end{eqnarray} 
The coefficients $f_i$ are constants. Higher orders in $r$ have been suppressed;
they are not restricted by regularity conditions.   
Recall that $\phi$ is the coordinate on the non-contractible cycle of the solid torus 
and therefore $d\phi$ is regular.
This means, in particular, that $f_{3}$ can be non-zero at the horizon.  
Expressing the $dr$ and $dt$ via \eqref{transxy} in terms of $dx$ and $dt$ and 
requiring the absence of $0/0$ expressions, reveals that for the spin-3
field to be regular at $r=0$ it must have the form 
\begin{equation}
ds_3^3 = f_5 ( r^2 dt^2 + dr^2 )d\phi + f_3 d\phi^3 
\end{equation} 
All solutions found in the literature so far have precisely this structure, namely
\begin{equation}
ds_3^3 = d\phi\times ds^2{(\hbox{black hole})}
\end{equation}  
even away from the origin. 

Note that the shift function, characterizing 
angular momentum, does not appear at leading order in the near horizon geometry. This 
issue affects the spin-2 and spin-3 fields. For the metric one expects a 
term $N^{\phi} dt d\phi $. For reasons explained above, $dt$ is singular and a necessary 
condition for regularity is that $N^\phi$ must vanish at the horizon. We thus see that 
imposing not only finiteness but also single-valuedness imply that $N^\phi$ must 
vanish at least quadratically at the horizon. This is of course the case for the BTZ 
metric, as can be verified directly and also for the Kerr solution.  

\section{General near horizon analysis}
\label{sec:holonomies}

We now analyze the general structure of solutions near the horizon. 
We turn on all charges. In particular, we do not set the charges in $a_\phi$ and 
$b_\phi$ equal, which would correspond to `non-rotating' solutions.\footnote{For 
background and notation see e.g. \cite{Banados:2012ue}.} 
This case was
mostly considered so far in the literature. 
An important lesson learned from the analysis of this section is that holonomy 
conditions imposed on gauge fields do not imply regularity on the metric-like fields.  
To be specific we consider $N=3$, but it will be clear that all our results 
are valid for any $N$. 

We consider solutions of $F_{\mu\nu}=0$
\begin{equation}
\begin{array}{ll}
A_t = g_{1}^{-1} a_t\, g_{1}  &  \ \ \ \ B_t = g_{2}^{-1} b_t\, g_{2} \\
A_\phi =  g_{1}^{-1} a_\phi\, g_{1}  & \ \ \ \ B_\phi =  g_{2}^{-1} b_\phi\, g_{2} \\
A_r =  g_1^{-1} \partial_r g_{1}  & \ \ \ \  B_r =  g_2^{-2} \partial_r g_{2}  
\label{A}
\end{array} 
\end{equation} 
where
\begin{equation}\label{eqns}
[a_t,a_\phi]=0 \ \ \ \ \ \ \   [b_t,b_\phi]=0
\end{equation} 

Here $g_1,g_2\in SL(3)$ are functions of $r$
and $a_t,\dots,b_\phi\in sl(3)$ are constant. 
From these two connections we form the dreibein $e_\mu ={1\over 2}(A_\mu - B_\mu)$ and 
the spin-2 and spin-3 fields~\cite{Stefan^3} 
\begin{eqnarray}\label{g2g3}
g_{\mu\nu}=\mbox{Tr}(e_\mu e_\nu)\qquad g_{\mu\nu\rho}
=\mbox{Tr}(e_{(\mu} e_\nu e_{\rho)})
\end{eqnarray} 

We seek group elements $g_1$ and $g_2$ 
and matrices $a_t,b_t,a_\phi,b_\phi$ such that $g_{\mu\nu}$ 
and $g_{\mu\nu\rho}$ are regular. 
The regularity conditions can be split into three classes, on which we elaborate below.  

\begin{enumerate}
\item {\bf Topological:} $a_t$ and $b_t$ must have trivial holonomies 
on the contractible cycle
\begin{equation}\label{holfin}
e^{\oint a_tdt} = e^{\oint b_tdt} = \mathbbm{1}  
\end{equation} 
\item {\bf Finiteness:} Since the 1-form $dt$ is singular, 
the time component of the vielbein $e_\mu dx^\mu$ must satisfy 
\begin{equation}\label{et=02}
e_t=0 \ \ \ \ \ \  \mbox{at} \ \ \ \ \ \  r=0 
\end{equation} 

\item
{\bf Single-valuedness:} As discussed above, in the vicinity of the horizon 
at $r=0$, the metric and spin-3 field must 
read\footnote{Recall that angular momentum ($dtd\phi$) only appears at ${\cal O}(r^2)$ 
at the horizon.}, 
\begin{equation}
\begin{aligned}\label{reg3}
ds_2^2 &=& f_1 (r^2 dt^2 +  dr^2) + f_2 d\phi^2  + \mbox{ (higher powers of }r)\\
ds_3^3 &=& \Big(f_3( r^2 dt^2 + dr^2) + f_4 d\phi^2 \Big) d\phi 
+ \mbox{(higher powers of }r) 
\end{aligned}
\end{equation} 
where $f_i$ are constants. All other components must vanish at $r=0$ at a 
sufficiently high order. Special attention will be given to whether holonomy conditions 
suffice to guarantee the Hawking conditions
\begin{equation}\label{Hawking}
{g_{tt} \over g_{rr}} = {g_{tt\phi} \over g_{rr\phi}} =r^2 
\end{equation} 
as required by regularity. 
  
\end{enumerate}

Let us study conditions 1., 2. and 3. one by one.

\subsection{Holonomy conditions and diagonal fields}
\label{sec:diagonal}

Eqn. (\ref{holfin}) means that $a_t$ and $b_t$ must have trivial holonomy and 
live in the Jordan class of diagonal matrices, with $i\times$(integer eigenvalues). 
This means 
there exists constant matrices $M_1, M_2$ such that \begin{equation}\label{atbt}
\begin{aligned}
a_t &=& M_1^{-1} a_t^{(d)} M_1 \ \ \  \ \ \  a_t^{(d)} =  i\left(\begin{array}{ccc}
p_1 & 0 & 0 \\ 
0 & p_2  & 0 \\ 
0 & 0 & -p_1-p_2
\end{array} \right) \\
\noalign{\vskip.2cm}
b_t &=& M_2^{-1} b_t^{(d)} M_2 \ \ \  \ \ \  b_t^{(d)} =  i\left(\begin{array}{ccc}
q_1 & 0 & 0 \\ 
0 & q_2  & 0 \\ 
0 & 0 & -q_1-q_2
\end{array} \right) 
\end{aligned}
\end{equation}    
where, to satisfy condition (\ref{holfin}), $p_1,p_2,q_1,q_2$ must be integers.  
We shall see below that further regularity conditions demand $q_1=p_1$ and $q_2=p_2$, 
up to trivial permutations of the eigenvalues.  

Note also \eqref{eqns}, i.e.  that $a_t$ commutes with $a_\phi$, and $b_t$ 
commutes with $b_\phi$. 
This implies that $M_1,M_2$ also make $a_\phi, b_\phi$ diagonal, 
that is 
\begin{equation}\label{apbp}
\begin{aligned}
a_\phi &=& M_1^{-1} a_\phi^{(d)} M_1 \qquad  a_\phi^{(d)} =  \left(\begin{array}{ccc}
-2 \lambda_1 & 0 & 0 \\ 
0 & \lambda_1 + \lambda_2 & 0 \\ 
0 & 0 & \lambda_1- \lambda_2
\end{array} \right) \\
\noalign{\vskip.2cm}
b_\phi &=& M_2^{-1} b_\phi^{(d)} M_2 \qquad  b_\phi^{(d)} =  \left(\begin{array}{ccc}
-2 \rho_1 & 0 & 0 \\ 
0 & \rho_1 + \rho_2 & 0 \\ 
0 & 0 & \rho_1- \rho_2
\end{array} \right) 
\end{aligned}
\end{equation}
The four charges $T_1,W_1,T_2,W_2$ are now parameterized in terms of the four
eigenvalues 
$\lambda_1,\lambda_2,\rho_1,\rho_2$. The exact relation is not necessary for this 
analysis. We shall come back to this point in Sec.~\ref{sec:GSH} where the above 
parametrization, which was introduced in \cite{deBoer:2013gz}, 
proves very convenient.  
     
Note that for generic values of the two charges $T$ and $W$ the three eigenvalues 
of $a_\phi$ are nondegenerate and as long as the discriminant of the 
characteristic equation is positive, i.e. 
$4\,T^3-27\,W^2>0$, they are 
all real. In this case the matrix $M_1$ which diagonalizes $a_\phi$ (and also $a_t$) 
is also real. The condition on the discriminant is also the condition for the 
black hole to be above extremality \cite{Maxetal}.     
   
\subsection{A finite vielbein on the torus: $e_t=0$ at $r=0$}  

The 1-form $dt$ is singular and must always appear multiplied by a function that 
vanishes at the horizon. This condition may seem natural and easy to implement. Note 
however, that as $N$ increases, the number of conditions grows rapidly. In the 
$sl(N)\times sl(N)$ theory, there are $N-1$ metric-like fields 
$g_{\mu\nu},g_{\mu\nu\rho},g_{\mu\nu\rho\sigma}$, etc. Each of these fields has components 
involving the 1-form $dt$ at various orders. To give some examples,  
\begin{eqnarray}\label{dt-components}
ds^2 &=&  \cdots + g_{tt}\, dt^2 + g_{t\phi}\, dt d\phi + \cdots\nonumber  \\
ds^3 &=& \cdots + g_{ttt}\, dt^3 + g_{tt\phi}\, dt^2 d\phi + \cdots  \\
ds^4 &=& \cdots + g_{tttt}\, dt^4  + g_{ttt\phi}\, dt^3 d\phi + 
g_{tt\phi\phi}\, dt^2 d\phi^2 +\cdots\nonumber
\end{eqnarray} 
where
\begin{equation}\label{greg}
g_{\mu_1....\mu_n} = \mbox{Tr} ( e_{(\mu_1} \cdots e_{\mu_n)} )
\end{equation}  
Imposing the necessary conditions to achieve regularity term by term may become quite tedious. Fortunately, the gauge formulation allows a solution that applies to all components at once.   

The idea is to implement regularity on the connections, $A_\mu, B_\mu$ 
rather than on the metric-like fields. But there is a subtlety which we now 
discuss. In principle, the connections $A_\mu(x)dx^\mu$ and $B_\mu(x)dx^\nu$ 
can be regular functions if they satisfy 
\begin{equation}
A_{t}(r) \rightarrow 0\qquad B_t(r)\rightarrow 0\qquad \hbox{as}\qquad r\rightarrow 0
\end{equation} 
(plus other conditions from single-valuedness).  
The trouble is that stationary 
and circularly symmetric solutions cannot 
fulfill this property. Recall
\begin{equation}
A_t(r) = g^{-1}(r)\,  a_t \, g(r) 
\end{equation} 
where $a_t$ is constant. Therefore, if $A_t(r_0)=0$ at any point $r_0$, 
then invertibility of $g$ implies $a_t=0$ and hence $A_t(r)=0$ at all points. 
The same holds for $B_t$. 

This obstruction has nothing to do with higher spin fields and shows 
up for $sl(N)$, including the very well-known case $N=2$.  
The static, circularly symmetric connections associated to BTZ black holes satisfy 
the following properties. The dreibein, the difference of both connections,  
\begin{equation}\label{emu}
e_\mu ={1 \over 2}( A_\mu - B_\mu)
\end{equation} 
is regular at the horizon, $e_t=0$ at $r=0$, but the spin connection
$$
w_\mu ={1 \over 2}(  A_\mu + B_\mu)
$$
is singular. This means that the metric field is regular but not the connection.  Of 
course the curvature is again regular. 
Therefore, as long as we are interested in regularity of the metric-like 
fields $g_{\mu_1\mu_2\dots}$, it suffices to impose  
\begin{equation}\label{et=0}
e_t(r) \rightarrow 0 \qquad\hbox{as}\qquad r\rightarrow 0.
\end{equation} 

One may wonder whether (\ref{et=0}) implies conditions that could over-determine the 
problem. This is not case. From (\ref{emu}) we see that (\ref{et=0}) implies 
\begin{equation}\label{et=02}
g_1^{-1} a_t g_1 - g_2^{-1} b_t g_2 =0 \qquad \Rightarrow \qquad  b_t = S^{-1} a_t S 
\end{equation} 
where $S:=g_1 g_2^{-1}$. In words, our regularity condition imply that $a_t$ and $b_t$ 
must be in the same $sl(N,\mathbb{R})$ class. 
We know already that $a_t$ and $b_t$ are both in the diagonal class because they must 
satisfy  (\ref{holfin}). Condition (\ref{et=02}) now implies that they must satisfy
\begin{equation}
\mbox{Tr}(a_t^2) = \mbox{Tr}(b_t^2)  \ \ \ \ \  \mbox{Tr}(a_t^3) = \mbox{Tr}(b_t^3) 
\end{equation} 
These conditions can be solved by setting
\begin{equation}
q_1 = p_1\qquad\hbox{and}\qquad q_2 = p_2
\end{equation}  
in (\ref{atbt}), that is
\begin{equation}\label{at=bt}
b_t^{(d)} = a_{t}^{(d)}
\end{equation}

\subsection{Single valued fields on the torus}
\label{sec:SV}

We now address condition (\ref{reg3}) and build fields with all desired properties at
the center of the solid torus. Collecting the information from the two previous 
paragraphs, we first note that the vielbein components
can be written as  
\begin{equation}
\begin{aligned}\label{aphi2}
e_t &= V^{-1}  \Big(  U^{-1} a^{(d)} _t U -  a_t^{(d)}  \Big) V \\
e_\phi &= V^{-1}  \Big(  U^{-1} a^{(d)} _\phi U -  b_\phi^{(d)}\Big) V  \\
e_r &= V^{-1} \Big(  U^{-1} \partial_r U    \Big) V 
\end{aligned}
\end{equation}
where $V=M_2\,g_2$ and $U = M_1^{\phantom{-1}}\!\!\!
g_1^{\phantom{-1}}\!\!\! g_2^{-1} M_2^{-1}$ (and we have used (\ref{at=bt})).  
Since the components 
$g_{\mu\nu},g_{\mu\nu\rho}$ are extracted from the  
traces Tr$(e^2)$ and Tr$(e^3)$, the matrix $V$ is irrelevant and can be set to 
one with no loss of generality. From now on we assume 
$$
V=\mathbbm{1} 
$$
The horizon (center of the solid torus) is defined as the point where $e_t=0$. 
Inspecting (\ref{aphi2}) we see that this requires that $U(r=0)$ must commute 
with $a_t^{(d)}$. 
Then, in the 
vicinity of the horizon we expand, without loss of generality, 
\begin{equation}\label{U}
U = \mathbbm{1} + r X_1 + r^2 X_2 + r^3 X_3 + \cdots
\end{equation} 
where $X_1,X_2,X_3$ are, at this point, general matrices, only constrained by 
the condition $\det U=1$.  

To implement conditions (\ref{reg3}) requires a bit more work, 
but is straightforward. We plug (\ref{U}) into (\ref{aphi2}), ignoring the 
factor $V$, and compute the 
traces (\ref{g2g3}) to first order in $r$.  This gives expressions for 
$g_{\mu\nu}$ and $g_{\mu\nu\rho}$. To impose all regularity conditions 
(\ref{reg3}) at $r=0$ we shall not need to go beyond the linear 
term in \eqref{U}, i.e. $X_2,\,X_3$ and the higher orders are not constrained by 
the regularity conditions at the horizon.  
To comply with det$(U)=1$ we set the matrix $X_1$ to be 
\begin{equation}\label{X1}
X_1 = \left(\begin{array}{ccc}
x_1 & x_2 & x_3 \\ 
x_4 & x_5 & x_6 \\ 
x_7 & x_8 & -x_1-x_5
\end{array} \right) \\
\end{equation}  

The general expressions obtained are long and not worth displaying here, 
but we show some components that should be zero, which implies conditions on $X_1$.  
For example,  
\begin{equation}
g_{r\phi} =2\,x_1(\lambda_2-\rho_2-3\,\lambda_1+3\,\rho_1)
+4\,x_5(\lambda_2-\rho_2) + {\cal O}(r) \,.\label{cond1}
\end{equation} 
From (\ref{reg3}) we know that this component must vanish at $r=0$. This is our first 
condition on the allowed group elements. One can set $g_{r\phi}=0$ to order 
${\cal O}(1)$ by solving (\ref{cond1}) with respect to $x_5$ 
\begin{equation}
x_5 = {3\,(\lambda_1-\rho_1)-(\lambda_2-\rho_2)  \over \lambda_2-\rho_2}
\,{x_1 \over 2}
\end{equation} 
No other conditions follow from (\ref{reg3}). 

We now move to the spin-3 field. First, one finds a non-zero value for 
\begin{equation}
g_{\phi\phi r} = 3\,x_1\big(3(\lambda_1-\rho_1)-(\lambda_2-\rho_2)\big)
\big(3(\lambda_1-\rho_1)+(\lambda_2-\rho_2)\big)+{\cal O}(r)\,. 
\end{equation} 
According to (\ref{reg3}) this component must vanish at $r=0$. 
The only solution, keeping all charges unconstrained is, 
\begin{equation}
x_1=0
\end{equation} 
Next, we note the components
\begin{equation}
\begin{aligned}\label{cgttt}
g_{ttt} &= -3  (p_1-p_2)(p_1+2p_2)(p_2+2p_1)(x_2 x_6 x_7- x_3 x_4 x_8)\, r^3 
+{\cal O}(r^4)\\
g_{ttr} &= -9(p_1^2+p_2^2+ p_1p_2) (x_2 x_6 x_7+ x_3 x_4 x_8)\,r^2
+{\cal O}(r^3)
\end{aligned}
\end{equation} 
which, as discussed before, should vanish at the order displayed. 
For generic integers 

$p_1,\,p_2$ this requires
\begin{equation}\label{condx}
x_2 x_6 x_7=0 \qquad\hbox{and}\qquad   x_3 x_4 x_8 =0 
\end{equation} 
and there are no further conditions from the regularity of $g_3$. 

The equations \eqref{condx} can be solved in various ways.  
For instance for $\{x_3=x_6=0\}$ and $\{x_7=x_8=0\}$ one finds 
\begin{eqnarray}\label{sol1}
{g_{tt} \over g_{rr}} = {g_{tt\phi} \over g_{rr\phi}} = (p_2-p_1)^2\,r^2+{\cal O}(r^3)
\end{eqnarray} 
For $\{x_2=x_8=0\}$ and $\{x_4=x_6=0\}$ one obtains 
\begin{eqnarray}\label{sol2}
{g_{tt} \over g_{rr}} = {g_{tt\phi} \over g_{rr\phi}} 
= (2\,p_1+p_2)^2\,r^2+{\cal O}(r^3) 
\end{eqnarray} 
and for $\{x_2=x_3=0\}$ and $\{x_4=x_7=0\}$
\begin{eqnarray}\label{sol3}
{g_{tt} \over g_{rr}} = {g_{tt\phi} \over g_{rr\phi}} 
= (2\,p_2+p_1)^2\,r^2+{\cal O}(r^3) 
\end{eqnarray} 
Clearly, they follow from each other by permutation of the eigenvalues of 
$a_t$ and $b_t$. The three remaining minimal solutions (with only one factor 
from each triplet in \eqref{condx} set to zero) give ratios $g_{tt}/g_{rr}$ and 
$g_{tt\phi}/g_{rr\phi}$ which (i) depend on some of the remaining non-zero $x's$ and 
(ii) are not equal. We will discuss one of them further in Section \ref{sec:SRS}. 

For the solutions displayed above regularity can be achieved in various ways: 
$(a)$ $p_2=p_1\pm1$, $(b)$ $p_2=-2\,p_1\pm1$ and $(c)$ $p_1=-2\,p_2\pm 1$. 
But clearly there are other possibilities which lead to a conical excess. 
Even for $p_1=1, p_2=0$, which is the case mostly discussed in the literature, 
there are two options, one leading to a regular solution and the 
other to a conical surplus by $2\pi$.  

\section{Solution of the holonomy conditions}
\label{sec:GSH}

So far we have parametrized the solutions of the holonomy conditions by 
two integer eigenvalues of $a_t$, $p_1$ and $p_2$, 
and  we have also parametrized $a_\phi$ by 
its two eigenvalues, $\lambda_1$ and $\lambda_2$. 
We have used that $a_t$ and $a_\phi$ can be simultaneously 
diagonalized, but we have not established relations between the eigenvalues 
and physical
quantities, i.e. charges and chemical 
potentials which enter the thermodynamic description developed in 
\cite{Banados:2012ue}.
When that description is used, the holonomy conditions are solved by expressing 
the charges in terms of the chemical potentials, which amounts to solving 
cubic equations (for $N=3$). To identify the physical charges in our solutions, 
we use, following \cite{Banados:2012ue}, 
the Casimirs of $a_\phi$. This allows  
for a generic solution of the holonomy conditions which expresses the 
chemical potentials through the charges which are parametrized by 
$\lambda_1$ and $\lambda_2$ 
and which is compatible with the analysis in \cite{Banados:2012ue}.

We will give a brief constructive description of how to arrive at this simple 
generic solution of the holonomy conditions.
Consider the usual principally embedded $sl(2) \rightarrow sl(3)$ gauge connections
in highest weight gauge\footnote{A similar analysis applies to the 
second $SL(3)$ factor, where we start with 
$b_\phi=\begin{pmatrix}0&1&0\\\bar T/2&0&1\\ \bar W&\bar T/2&0\end{pmatrix}$.}
\begin{eqnarray}\label{prin}
a_\phi &=&  \left(\begin{array}{ccc}
0 & T/2 & W \\ 
1 & 0 & T/2 \\ 
0 & 1 & 0
\end{array} \right)
\end{eqnarray}
and\footnote{As discussed in \cite{Valdivia}, it is convenient to introduce chemical
potentials in the time component of the fields. In this way the time evolution
generates an `allowed gauge transformation'.} 
\begin{eqnarray}\label{at}
a_t &=& i\,\mu\, a_\phi
+ i\,\nu \left(a_\phi^2-{1\over3}\mbox{Tr}(a_\phi^2)\mathbbm{1}\right)
\end{eqnarray} 
$T$ and $W$ are the charges and $\mu$ and $\nu$ the chemical 
potentials. Using \eqref{apbp} we can parametrize the charges
in terms of the $a_{\phi}$ eigenvalues as
\begin{equation}
\begin{aligned}\label{TWcas}
T&= {1\over 2}{\rm Tr}(a_{\phi}^{2})
=3\,{{\lambda_1}}^{2}+{{\lambda_2}}^{2}\\
W&= {1\over 3}{\rm Tr}(a_{\phi}^{3})
=-2\,{\lambda_1}\, \left( {\lambda_1}
-{\lambda_2} \right) 
\left( {\lambda_1}+{\lambda_2} \right)
\end{aligned}
\end{equation}
If we now combine $a_{t}$ in \eqref{prin} with \eqref{atbt} and  \eqref{apbp}, 
we can solve for $\mu,\,\nu$ in terms of $\lambda_1,\,\lambda_2$:  
\begin{equation}
\begin{aligned}\label{holgenmu}
\mu_{(p_1,p_2)}&=&{1\over 2}\,{\frac { \left( 3\,{{\lambda_1}}^{2}-6\,{\lambda_1}\,
{\lambda_2}-{{\lambda_2}}^{2} \right) {p_1}}{{\lambda_2}\, \left( 3\,{\lambda_1}
-{\lambda_2} \right)  
\left( 3\,{\lambda_1}+{\lambda_2} \right) }}
+{\frac { \left( 3\,{{\lambda_1}}^{2}
-{{\lambda_2}}^{2} \right) {p_2}}{{\lambda_2}\, 
\left(3\,{\lambda_1}-{\lambda_2} \right)  \left( 3\,{\lambda_1}
+{\lambda_2} \right) }}\\
\nu_{(p_1,p_2)}&=&{3 \over 2}\,{\frac { \left( {\lambda_1}
+{\lambda_2} \right) {p_1}}{{\lambda_2}\, \left( 3\,{\lambda_1}-{
\lambda_2} \right)  \left( 3\,{\lambda_1}+{\lambda_2} \right) }}
+{\frac {3 {\lambda_1}\,{p_2}}{{\lambda_2}\, \left( 3\,{\lambda_1}
-{\lambda_2} \right)  \left( 3\,{\lambda_1}
+{\lambda_2}\right) }}
\end{aligned}
\end{equation}
It is worth stressing that these expressions for the chemical potentials 
are the general solutions of the holonomy conditions. They are labeled 
by the two integers $p_1\,,p_2$ which 
parameterize all possible holonomy branches, while 
$\lambda_1,\lambda_2$ parametrizes the charges.

The above result is purely algebraic. We will now demonstrate 
its consistency with the thermodynamic analysis of \cite{Banados:2012ue}. 
From \eqref{TWcas} it is clear that the eigenvalues are implicitly 
defined in terms of the charges, i.e. $\lambda_1(\,T,\,W),\lambda_2(\,T,\,W)$. 
Consider 
\begin{equation}
G[T,W]_{(p_1,p_2)}=-3\,{p_1}\,{\lambda_1}+(p_1+2\,{p_2})\,{\lambda_2}
\end{equation}

Using the implicit definitions in \eqref{TWcas}, one finds
\begin{equation}
\begin{aligned}\label{munu}
\mu &\equiv {\partial G[T,W] \over \partial T}\cr
&={\partial G \over \partial \lambda_i}{\partial \lambda_{i} \over \partial T}
={1\over 2}\,{\frac { \left( 3\,{{\lambda_1}}^{2}-6\,{\lambda_1}\,{\lambda_2}
-{{\lambda_2}}^{2} \right) {p_1}}{{\lambda_2}\, \left( 3\,{\lambda_1}
-{\lambda_2} \right)  \left( 3\,{\lambda_1}+{\lambda_2} \right) }}
+{\frac { \left( 3\,{{\lambda_1}}^{2}
-{{\lambda_2}}^{2}\right){p_2}}{{\lambda_2}\, \left( 3\,{\lambda_1}-{\lambda_2} \right)  
\left( 3\,{\lambda_1}+{\lambda_2} \right) }}\cr
\noalign{\vskip.2cm}
\nu &\equiv {\partial G[T,W] \over \partial W}\cr 
&={\partial G \over \partial \lambda_i}{\partial \lambda_{i} \over \partial W}
={3 \over 2}\,{\frac { \left( {\lambda_1}
+{\lambda_2} \right) {p_1}}{{\lambda_2}\, \left( 3\,{\lambda_1}
-{\lambda_2} \right)  \left( 3\,{\lambda_1}+{\lambda_2} \right) }}
+{\frac {3 {\lambda_1}\,{p_2}}{{\lambda_2}\, \left( 3\,{\lambda_1}-{\lambda_2} \right)  
\left( 3\,{\lambda_1}+{\lambda_2}\right) }} 
\end{aligned}
\end{equation}
in agreement with \eqref{holgenmu}.
In fact, if \eqref{munu} are satisfied, then the eigenvalues of $a_t$ are
\begin{equation}\label{eig}
\mbox{Eig}(a_t) = 2\,\pi\, i\left(p_1,p_2,-p_1-p_2\right)
\end{equation} 
and hence the holonomy condition (\ref{holfin}) holds.

Analogously, let  $F[\mu,\nu]_{(p_1,p_2)}$ be the following implicit 
function of the chemical potentials, 
\begin{eqnarray}\label{F}
F[\mu,\nu]_{(p_1,p_2)} &=& {1\over 2}\,{\frac { \left( 3\,{{\lambda_1}}^{4}
+30\,{{\lambda_1}}^{3}{
\lambda_2}-6\,{\lambda_1}\,{{\lambda_2}}^{3}+{{\lambda_2}}^{4}
-12\,{{\lambda_1}}^{2}{{\lambda_2}}^{2} \right) {p_1}}{{\lambda_2}\, 
\left( 3\,{\lambda_1}-{\lambda_2} \right)  \left( 3\,{
\lambda_1}+{\lambda_2} \right) }}\\
\noalign{\vskip.2cm}
&&\qquad\qquad+{\frac { \left( 3\,{{\lambda_1}}^{4}
+{{\lambda_2}}^{4}-12\,{{\lambda_1}}^{2}{{\lambda_2}}^{2}\right) 
{p_2}}{{\lambda_2}\, \left( 3\,{\lambda_1}
-{\lambda_2} \right)  \left( 3\,{\lambda_1}+{\lambda_2} \right) }}
\end{eqnarray}
where it is now understood to use equations \eqref{holgenmu} to implicitly define the 
eigenvalues in terms of the chemical potentials, i.e., $\lambda_1(\mu,\nu),
\lambda_2(\mu,\nu)$. 
It then follows from \eqref{TWcas}
\begin{equation}
\begin{aligned}\label{TW}
T &= {\partial F[\mu,\nu] \over \partial \mu}
={\partial F \over \partial \lambda_i} {\partial \lambda_{i} \over \partial \mu}
= 3\,{{\lambda_1}}^{2}
+{{\lambda_2}}^{2}\\ 
W &= {\partial F[\mu,\nu] \over \partial \nu}
={\partial F \over \partial \lambda_i} {\partial \lambda_{i} \over \partial \nu}
=-2\,{\lambda_1}\, \left( {\lambda_1}-{\lambda_2} \right) 
\left( {\lambda_1}+{\lambda_2} \right)
\end{aligned}
\end{equation} 
in agreement with \eqref{TWcas}.
It goes without saying that $F[\mu,\nu]$ and $G[T,W]$ are, respectively, 
the canonical and micro-canonical actions that can be computed 
with the methods discussed in \cite{Banados:2012ue}. Here our goal is to study 
properties of the solutions and we refer the interested 
reader to this reference for more information on how to derive 
$F[\mu,\nu]$ and $G[T,W]$ from the Chern-Simons action. 

If we impose in addition to the holonomy condition that the solution has a BTZ limit 
in which $W \rightarrow 0$ as $\nu \rightarrow 0$, then we see from  
\eqref{munu} and \eqref{TW} that $\nu=W=0$ leads to
\begin{equation}\label{BTZlimit}
0={\lambda_1}\, \left( {\lambda_1}-{\lambda_2}\right)  
\left( {\lambda_1}+{\lambda_2} \right) \,\,\,,\,\,\,0
= \left( {\lambda_1}+{\lambda_2} \right) {p_1}+2\,{\lambda_1}\,{p_2}
\end{equation}
This only possesses solutions 
with one unconstrained charge or, equivalently, one unconstrained 
$\lambda_{i}$ eigenvalue, if one of the integer eigenvalues of $a_t$, 
$p_1,p_2$ or $p_1+p_2$, is zero.

\section{A simple exact regular solution}
\label{sec:SRS}

We now turn to the construction of a simple exact regular solution 
which has a BTZ limit. As stated in 
section \ref{sec:GSH}, such solutions must belongs to the 
class of $a_t$ matrices with at least one eigenvalue equal to zero.
In particular, we will use the class with $p_1=1,p_2=0$, i.e., with eigenvalues 
$(2 \pi , 0, -2\pi)$. This is the class which has mostly been considered in the 
literature, starting from \cite{GutperleKraus}. 
Our exact solution will be constructed using what we will call the `minimal 
prescription', which is given by 
\begin{equation}\label{minimal}
U=e^{r X_1}
\end{equation}
where $X_1$ is the matrix \eqref{X1}, partially determined by the single-valuedness 
conditions on the solid torus. 
It is `minimal' in the sense that the group element is determined by the same 
algebra element $X_1$ at all orders in $r$.  
As explained in Section \ref{sec:SV}, 
we need to set $x_1=x_5=0$. Then, for the particular values $p_1=1,p_2=0$, 
besides the branches shown in \eqref{sol1} and \eqref{sol3} which are 
valid for generic $p_i$'s, 
\footnote{On the branches \eqref{sol1} and \eqref{sol3} the minimal description 
does not lead to field configurations with four independent parameters. 
A non-minimal 
description might, but a systematic analysis is involved and beyond the 
scope of this note.}, we can choose 
$x_3=x_7=0$. This solves the equations \eqref{condx} and gives 
\begin{eqnarray}\label{sol4}
{g_{tt} \over g_{rr}} = {g_{tt\phi} \over g_{rr\phi}} = r^2+{\cal O}(r^3)
\end{eqnarray} 
There are no further conditions from regularity at the horizon 
and therefore the parameters $x_2,x_4,x_6,x_8$ are left free. However, 
if we then construct the solution using \eqref{minimal}, we 
find that these parameters enter the metric fields only through the two products 
$x_2 x_4$ and $x_6 x_8$ and that the metric fields are of the general form  
\begin{equation}
\begin{aligned}\label{metf}
ds^2=&g_{rr}\,dr^{2}+g_{tt}\,dt^2+g_{t\phi}\,dt\,d\phi+g_{\phi\phi}\,d\phi^2\\
ds^3=&d\phi \times (g_{rr\phi}\,dr^{2}+g_{tt\phi}\,dt^2+g_{t\phi\phi}\,dt\,d\phi
+g_{\phi\phi\phi}\,d\phi^2)\\
\end{aligned}
\end{equation}
All other components vanish. Furthermore, 
the component $g_{rr}$ is always a constant\footnote{It is worth mentioning 
that with the minimal prescription the exact form of the metric fields is of the form 
\eqref{metf}, independent of the branch.} and we can 
achieve $g_{rr}=1$ either by rescaling $r$ or by the choice 
\begin{equation}
x_6\,x_8=4-{x_2}\,{x_4}        
\end{equation}
If we then demand that the metric fields are described by four 
independent charges and, at the same time, the existence of a BTZ 
limit in $ds^2$, we need to set $x_2\,x_4=2$. 

This being done, the non-vanishing components of the metric are\footnote{The factors of 
$i$ in the components with an odd power of $dt$ are due to our use of 
Euclidean signature.}
\begin{equation}
\begin{aligned}\label{spin2}
g_{rr}=&1\cr
g_{tt}=&\sinh^{2}(r)\cr
g_{t\phi}=&{i\over 2}\,\left( -{\rho_2}+3\,{\rho_1}-{\lambda_2}
+3\,{\lambda_1}\right)\sinh^{2}(r)\cr
g_{\phi\phi}=&{1\over 4}\, \left( 3\,{\rho_1}
-{\rho_2} \right)  \left({\lambda_2}
-3\,{\lambda_1} \right) \sinh^{2}(r) 
-{\frac {9}{16}}\,\left( {\rho_1}+{\rho_2} \right)  \left( {\lambda_1}
+{\lambda_2} \right) \sinh^{2}(2\,r) \\
&+{1\over 4}\, \left( {\lambda_2}
-{\rho_2} \right) ^{2}+{3\over 4}\, \left( {\lambda_1}-{\rho_1}
\right) ^{2}
\end{aligned}
\end{equation}
while the spin-3 field is 
\begin{equation}
\begin{aligned}\label{spin3}
g_{rr\phi}=&{1\over 8}\,\left({\lambda_1}+\,{\lambda_2}-\,{\rho_1}-\,{\rho_2}\right)\cr
g_{tt\phi}=&{1\over 32}\, \left( {\lambda_1}+{\lambda_2}-{\rho_1}-{\rho_2}\right)  
\left(3\,\sinh^{2}(2\,r)-8\,\sinh^{2}(r) \right)\cr
g_{t\phi\phi}=&{i\over 8}\, \left( {\lambda_1}+{\lambda_2}-{\rho_1}-{\rho_2}\right)  
\left( {\rho_2}-3\,{\rho_1}+{\lambda_2}-3\,{\lambda_1} \right) \sinh^{2}(r)\cr
&\qquad\qquad
-{3\,i\over 8}\,\left({\lambda_1}\,{\rho_2}-{\lambda_2}\,{\rho_1}\right)\sinh^{2}(2\,r)
\cr
g_{\phi\phi\phi}=&{3\over 64}\,\Big(\lambda_2\,\rho_2^2
+3\,\lambda_1^2\rho_2-3\,{\lambda_1}\,\rho_1^2
-\lambda_2^2\rho_2-\lambda_2^2 \rho_1+3\,\lambda_1^2 \rho_1
+\lambda_1\rho_2^2\cr
&+\,6\,{\lambda_2}\,{\rho_1}\,{\rho_2}
-6\,{\lambda_1}\,{\lambda_2}\,{\rho_2}
-6\,{\lambda_1}\,{\lambda_2}\,{\rho_1}
+6\,{\lambda_1}\,{\rho_1}\,{\rho_2}
-3\,{\lambda_2}\,{{\rho_1}}^{2}\Big)\sinh^{2}(2\,r)\cr
&-{1\over 16}\,\left(3\,{\rho_1}-{\rho_2} \right)\left( {\lambda_2}
-3\,{\lambda_1} \right)  \left( {\lambda_1}+{\lambda_2}
-{\rho_1}-{\rho_2} \right)\sinh^{2}(r)\cr
&-{1\over 8}\, \left( {\lambda_1}-{\rho_1} \right)  \left( -{\lambda_2}
+{\lambda_1}-{\rho_1}+{\rho_2} \right)  \left( {\lambda_1}
+{\lambda_2}-{\rho_1}-{\rho_2} \right) 
\end{aligned}
\end{equation}
In the BTZ limit, where $W_1 \rightarrow 0$ as $\nu_1 \rightarrow 0$ and 
$W_2 \rightarrow 0$ as $\nu_2 \rightarrow 0$, the spin-3 field is zero. 
As discussed at the end of section \ref{sec:GSH}, for this particular case, this 
happens when $\lambda_1 \rightarrow -\lambda_2$ and $\rho_1 \rightarrow -\rho_2$ 
at the same time.

The above solution is written in a proper radial coordinate such that 
$g_{rr}=1$. Often the use of Schwarzschild-like coordinates $r \rightarrow \ln(r)$,
where the horizon is at $r=1$, 
is more convenient. In these coordinates the asymptotic expansion 
(as $r \rightarrow \infty$) is   
\begin{equation}
\begin{aligned}\label{asymp}
g_{rr}=&{1\over r^2}\cr
g_{tt}=&{1\over 4}r^2+\mathcal{O}(1)\cr
g_{t \phi}=&{1\over 8}\,i \left( -{\rho_2}+3\,{\rho_1}-{\lambda_2}
+3\,{\lambda_1}\right) {r}^{2}+\mathcal{O}(1)\cr
g_{\phi \phi}=&-{9\over 64}\, \left( {\rho_1}+{\rho_2} \right)  \left( {\lambda_1}
+{\lambda_2} \right) {r}^{4}+\mathcal{O}(r^2)
\end{aligned}
\end{equation}
At large $r$ the leading term ($\sim r^4$) of $g_{\phi\phi}$  
vanishes in the BTZ limit,
i.e. these solutions are not small deformations of the BTZ 
black hole or, in other words, the limit is not smooth. This is a general feature and will be discussed in the next section.

\section{A ``no go theorem": asymptotics vs regularity}
So far we have concentrated on the near horizon 
properties, and we have imposed conditions on the group element $U$ to achieve 
regularity for a wide range of possibilities.  A natural question which now arises 
is whether the group element $U$ can be extended across the manifold, maintaining  
stationary and circularly symmetric fields such that the solution is 
asymptotically AdS? The answer is no, as we shall show. 
 
By asymptotically AdS we mean solutions where the asymptotic form of the metric 
(in Schwarzschild-like coordinates) is
\begin{equation}
ds^2 = ds^2_{\mbox{{\tiny AdS}}} + {1 \over r}(\mbox{corrections}) 
\end{equation} 
Most importantly, the AdS radius of 
$ds^2_{\mbox{{\tiny AdS}}}$ must not depend on the 
charges. 
This last requirement cannot be fulfilled when the spin-3 charge 
is active.

It is important to mention that this problem only arises when we impose the 
asymptotic condition at infinity  and, at the same time, require regularity
in the interior. 
Asymptotically AdS solutions which carry all charges can of course be constructed 
(see \cite{CampoleoniHenneaux,HenneauxRey,Stefan^3}). 
The trouble arises when, in addition, we impose 
regularity at the horizon. 
If one does so, the spin-3 source 
appears at the leading term in the metric and the value of the 
cosmological constants jumps when this source is turned on/off.
This phenomenon was 
first observed by Gutperle and Kraus in \cite{GutperleKraus}, 
for a particular choice of the $r$-dependent group element.
In \cite{GutperleKraus} this was interpreted 
holographically as the effect of turning on an irrelevant operator in the 
UV theory. 
We prove here that there exists no choice of group element such that the 
cosmological constant is stable\footnote{This problem 
does not arise if one works in the so called diagonal embedding where no spin-3 charge 
is present.}.    
Again we will restrict to $N=3$. 

A distinguished family of solutions to \eqref{eqns} are the (anti-) chiral fields
\begin{equation}
a_t=a_\phi\qquad\qquad\hbox{and}\qquad\qquad b_t=- b_\phi
\end{equation} 
where $a_\phi$ and $b_\phi$ as in Section \ref{sec:GSH} and 
with the group element
\begin{equation}\label{g1g20}
g_1 = g_2^{-1} = \left( \begin{array}{ccc}
r & 0 & 0 \\ 
0 & 1 & 0 \\ 
0 & 0 & {1 \over r}
\end{array}  \right) 
\end{equation}  

These solutions belong to the principal embedding $sl(2)\hookrightarrow sl(3)$
and carry all desired $W_3$ 
features. 

The spacetime connections, built as in (\ref{A}), have the form  
\begin{eqnarray}
A_\phi &=& \left( \begin{array}{ccc}
0 & T/(2 r) & W/r^2 \\ 
r & 0 & T/(2 r) \\ 
0 & r & 0
\end{array}  \right) = A_t \label{hp}\\
&=& \mbox{background + fluctuations (subleading as } r\rightarrow \infty)
\end{eqnarray} 
They have the desired structure: a charge-independent background + subleading 
terms carrying 
the charges.  This solution cannot be extended, however, all the way to the horizon.

Regularity at the center of the solid torus requires the holonomy conditions to be 
satisfied and thus the sources, chemical potentials, have to be turned on. 
Given $a_\phi$, the general solution to (\ref{eqns}) is
$a_t$ as given in \eqref{at}.
 
When $\nu\neq 0 $, the matrix $a_t$ in (\ref{at}) is no longer proportional to 
$a_\phi$. In particular, it has non-zero components in all of its

entries\footnote{here we use Lorentzian signature, hence there are no factors of $i$.} 
\begin{equation}\label{a}
a_t \simeq \left( \begin{array}{ccc}
a_{11} & a_{12} & a_{13} \\ 
\mu & a_{22} & a_{23} \\ 
\nu & \mu & a_{33}
\end{array}  \right)
\end{equation} 
Here, we have explicitly displayed the values for the lower left corner because they 
are the relevant ones for our analysis, and also the simplest ones. 
All other coefficients 
$a_{ij}$ are functions of $T,W,\mu$ and $\nu$. 

Given $a_t$ in (\ref{a}), the spacetime field $A_t$, using the group element 
displayed in (\ref{g1g20}), is 
\begin{equation}
A_t =  \left( \begin{array}{ccc}
a_{11} & {a_{12}/r} & {a_{13}/r^2} \\ 
\mu\, r & a_{22} & {a_{23}/r} \\ 
\nu\, r^2  & \mu\, r  & a_{33}
\end{array}  \right)\label{tr}
\end{equation} 
For large $r$, the components in the upper triangle become `fluctuations'. 
The components in the lower triangle grow with $r$. We clearly see that the 
spin-2 source $\mu$ appears at order $r$ while the spin-3 source $\nu$ appears 
at order $r^2$. This is the effect observed in \cite{GutperleKraus}. 

If one turns $\nu$ on/off, the asymptotics of the metric changes and the cosmological 
constant jumps by a factor of four, i.e. the radius of the asymptotic $AdS_3$ 
decreases by a factor of two.

A natural question is whether the choice (\ref{g1g20}) is mandatory. Perhaps a more 
general group element can avoid the term $\nu\, r^2$ in the lower corner? 
The question we ask is whether one can build solutions that are simultaneously 
regular at the horizon and which have the desired asymptotics.  
This is not the case, as we now prove. 

Consider a general group element $g(r)\in SL(3)$
\begin{equation}
g = \left( \begin{array}{ccc}
\alpha_1(r) & \alpha_2(r) & \alpha_3(r) \\ 
\alpha_4(r) & \alpha_5(r) & \alpha_6(r) \\ 
\alpha_7(r) & \alpha_8(r) & \alpha_9(r)
\end{array}  \right)
\end{equation} 
(One of these functions is fixed by the requirement det$(g_1)=1$, but we shall 
not need to impose this  explicitly.) We write again a general solution $a_t$ 
as in (\ref{at}). 

Given $g(r)$, $A_t,A_\phi$ are uniquely defined by (\ref{A}). 
It is useful to give names to some of their components, 
\begin{equation}
 A_\phi = \left( \begin{array}{ccc}
\star & \star & \star  \\ 
f_2(r) & X(r) & \star \\ 
o_2(r) & f_1(r) & \star
 \end{array} \right)  \ \ \ \ \ \ \  
 A_t = \left( \begin{array}{ccc}
\star & \star & \star \\ 
f_{4}(r) & \star  & \star \\ 
o_1(r) & f_3(r) & \star 
\end{array} \right)  \label{aa}
\end{equation}   
The symbol $\star$ represents functions which do not enter the argument. 
We  would like to have solutions approaching (\ref{hp}) at infinity, 
thus we need to find functions $\alpha_i(r)$ such that, asymptotically,
\begin{equation}\label{asymc1}
o_1(r),o_2(r),X(r) \rightarrow 0 
\end{equation} 
together with  
\begin{equation}\label{asymc2}
f_1(r),f_2(r) \rightarrow r 
\end{equation} 
It turns out that conditions (\ref{asymc1}) and (\ref{asymc2}) are inconsistent. 

To show this we invert the problem and write the functions $\alpha_i(r)$ 
in the group element in terms of $f_{1},f_2,f_3,f_4,o_1,o_2$ and $X$. 
Although long, it is easy to show that 
$\alpha_2(r), \alpha_5(r),\alpha_6(r),\alpha_8(r),\alpha_9(r)$ can all   
be expressed as functions of $f_1(r),f_2(r)$, $f_3(r)$, $o_1(r),o_2(r)$. 
The explicit formulae are emposing and not useful, so we don't display them here. 
The important point is that the following relation follows, 
\begin{equation}\label{X}
\nu \Big(f_1(r)f_2(r)-o_2 (r)X(r)\Big)  + \mu\, o_2(r)-o_1(r)  =0
\end{equation} 
In other words, no matter what the group element $g(r)$ is, the components of 
$A_\phi,A_t$ are related by (\ref{X}).  
 
The obstruction is now clear. Conditions (\ref{asymc1}) demand $o_1,o_2,X$ to vanish 
while (\ref{asymc2}) demand $f_1,f_2$ to diverge. But this is inconsistent 
with (\ref{X}). If the spin-3 source is set to zero, then $\nu=0$ and $o_1$ and $o_2$ 
can be set to zero. 

In conclusion, for generic values of the 
charges $T$ and $W$, regularity of the horizon requires $\nu \neq 0$, and this is 
inconsistent with a field asymptotically AdS of the form (\ref{hp}). 

\section{Acknowledgements}

MB is partially supported by FONDECYT Chile, grant \# 1141221. MB would also like 
to thank H. Nicolai and the Max Planck Institute for Gravitational Physics,
Potsdam-Golm, for hospitality. We would like to thank A. Campoleoni, S. Fredenhagen,
Wei Li and I.~Reyes for helpful discussions.

\end{document}